\documentclass{mn2e} 
 
\usepackage{amsfonts} 
\usepackage{amsmath} 
\usepackage{graphicx} 
 
\title[Strong clustering of underdense regions] 
{Strong clustering of underdense regions and the environmental  
 dependence of clustering from Gaussian initial conditions} 
 
\author[U. Abbas and R. K. Sheth] 
{Ummi Abbas$^{1}$ and Ravi K. Sheth$^{2}$\thanks{E-mail: shethrk@physics.upenn.edu (RKS)}\\ 
$^{1}$Laboratoire D'Astrophysique de Marseille,  
      Traverse du Siphon, B.P. 8, 13376 Marseille Cedex 12, France\\ 
$^{2}$Department of Physics \& Astronomy, University of Pennsylvania,  
      209 S. 33rd St., Philadelphia, PA 19104, USA}

\begin{document} 
\pagerange{\pageref{firstpage}--\pageref{lastpage}} 
 
\maketitle 
 
\label{firstpage} 
 
\begin{abstract} 
We discuss two slightly counter-intuitive findings about the  
environmental dependence of clustering in the Sloan Digital Sky  
Survey.  First, we find that  the relation between  
clustering strength and density is {\em not} monotonic:   
galaxies in the densest regions are more strongly clustered  
than are galaxies in regions of moderate overdensity;   
galaxies in moderate overdensities are more strongly clustered  
than are those in moderate underdensities;  
but galaxies in moderate underdensities are {\em less} clustered  
than galaxies in the least dense regions.   
We argue that this is natural if clustering evolved gravitationally  
from a Gaussian field, since the highest peaks and lowest troughs  
in Gaussian fields are similarly clustered.   
The precise non-monotonic dependence of galaxy clustering on density  
is very well reproduced in a mock catalog which is based on a  
halo-model decomposition of galaxy clustering.   
In the mock catalog, halos of different masses are all about 200  
times denser than the critical density, and the dependence of small  
scale clustering on environment is entirely a consequence of the  
fact that the halo mass function in dense regions is  
top-heavy---another natural prediction of clustering from Gaussian  
initial conditions.   
%That the mock is able to reproduce the SDSS measurements is strong  
%evidence that these assumptions are accurate.   
 
Second, the distribution of galaxy counts in our sample is rather  
well described by a Poisson cluster model.   
We show that, despite their Poisson nature, correlations with  
environment {\em are} expected in such models.   
More remarkably, the expected trends are very like those in standard  
models of halo bias, despite the fact that correlations with  
environment in these models arise purely from the fact that dense  
regions are dense because they happen to host more massive halos.   
This is in contrast to the usual analysis which assumes that it is  
the large scale environment which determines the halo mass function.   
\end{abstract} 
 
%\shortauthor{U. Abbas \& R. K. Sheth} 
%\keyphrases{Large scale structure, clustering} 
 
\begin{keywords} 
methods: analytical - galaxies: formation - galaxies: haloes - 
dark matter - large scale structure of the universe  
\end{keywords} 
 
\section{\protect\bigskip Introduction} 
This is the third in a series of papers which study the  
environmental dependence of galaxy clustering.   
In hierarchical models, there is a correlation between  
fluctuations on different scales.  This induces correlations  
between halo mass and/or formation and the larger scale environment  
of a halo (Mo \& White 1996; Sheth \& Tormen 2002) which, in turn,  
induce correlations between galaxies and their environments  
(Sheth, Abbas \& Skibba 2004; Abbas \& Sheth 2005).   
Abbas \& Sheth (2006) showed that halo bias---the correlation between  
halo mass and environment---was able to account for the environmental  
dependence of clustering in the Sloan Digital Sky Survey (hereafter  
SDSS).  In that study, a galaxy's environment was defined as the  
number of galaxies within $8h^{-1}$Mpc, and only a relatively small  
range of environments were considered:  the two-point correlation  
function $\xi(r|\delta)$ of galaxies in the densest third of the  
sample was shown to be about five times larger than that of the full  
sample, whereas the galaxies in the least dense 30\% were less  
strongly clustered on scales larger than about $0.1h^{-1}$Mpc.   
 
In this paper we show that clustering strength is not a monotonic  
function of environment in the least dense regions:  compared to  
the objects in the least dense 30\% of the sample, the galaxies  
in the least dense 10\% are {\em more} strongly clustered.   
Nevertheless, the statistical halo-bias based effect accounts  
very well for the observed non-monotonic relation between  
environment and clustering strength.   
Section~\ref{xidelta} presents our measurements, shows that a  
halo-model (see Cooray \& Sheth 2002 for a review) based mock  
catalog exhibits the same features as seen in the data.   
Section~\ref{nlbias} discusses a simple model of the effect,  
and a final section summarizes our results and discusses some  
implications.    
 
An Appendix discusses a somewhat surprising interpretation  
of the origin of halo bias.  This is motivated by the fact that  
two of the distributions which have routinely been found to provide  
a good description of galaxy counts in cells are the  
Thermodynamic or Generalized Poisson distribution  
(Saslaw \& Hamilton 1984; Sheth 1995), and the Negative Binomial  
distribution (Moran 1984).  These distributions provide a good  
description of the counts in our catalog as well; they are both  
examples of Poisson cluster models (Daley \& Vere-Jones 2003).   
We show that, despite their Poisson nature, Poisson cluster  
models are expected to show environmental effects.   
However, in such models, dense regions are dense because they  
happen to host massive halos.  The standard analysis of halo bias  
assumes that the large scale environment determines the halo mass  
function, rather than the other way around.   
Nevertheless, we show that the expected halo bias in Poisson  
cluster models bears surprising similarity to the standard models  
of halo bias (Mo \& White 1996; Sheth \& Tormen 2002).   
  
Throughout, we show results for a flat $\Lambda$CDM model for which  
$(\Omega_{0},h,\sigma_{8}) = (0.3,0.7,0.9)$ at $z=0$.   
Here $\Omega _{0}$ is the density in units of critical density  
today, $h$ is the Hubble constant today in units of  
$100$~km~s$^{-1}$~Mpc$^{-1}$,  
and $\sigma_{8}$ describes the rms fluctuations of the initial  
field, evolved to the present time using linear theory,  
when smoothed with a tophat filter of radius $8h^{-1}$~Mpc.   
The Very Large Simulation (VLS) we use to construct mock catalogs  
was made available to the public by the Virgo consortium.   
It was run with the same $\Lambda$CDM cosmology, and followed the  
evolution of $512^3$ particles in a cubic box with sides  
$L = 479h^{-1}$Mpc (Yoshida et al. 2001).   
Dark matter halos were identified in this particle distribution  
using the Friends-of-Friends method.  
Each halo has a mass, a position and a velocity.

\section{The environmental dependence of $\xi$}\label{xidelta} 
To study the environmental dependence of clustering, we began  
with a parent galaxy catalog drawn from a parent catalog which  
was slightly larger than the SDSS DR4 database, and volume limited  
to $M_r<-19.5$.   
This catalog contains about $78,000$ galaxies with accurate  
angular positions and redshifts; the associated comoving number  
density is $0.01~(h^{-1}$Mpc)$^{-3}$.   
 
The environment of each galaxy in this catalog was defined as  
the number of such galaxies $N_8$ within $8h^{-1}$Mpc.   
No attempt was made to correct for redshift space distortions,  
which, on these scales, should be relatively small (though not  
negligible; see discussion of Fig.~\ref{NgalMhalo} below).   
Figure~\ref{sdsspdf} shows the distribution of densities which  
results.  We have constructed four subsamples of this catalog  
on the basis of environment as follows.  The lowest density  
environments we probe use the 10\% of the objects with the  
fewest neighbours within $8h^{-1}$Mpc.  A slightly less severe  
cut uses 30\% rather than 10\% of the objects.   
We then do the same for overdense regions:  we select subsets  
containing 10\% and 30\% of the objects having the most neighbours  
within $8h^{-1}$Mpc.  Hashed regions indicate the various density  
thresholds which these cuts imply.  The actual overdensity thresholds  
$\delta_8 \equiv N_8/\langle N_8\rangle - 1$ are indicated in the  
upper right corner of the figure, with some abuse of notation:   
$\delta_{n}$ means $n\%$ of the galaxies were in lower density  
environments (i.e., $p(\le\delta_{n\%}) = n/100$).   
In what follows, we use these limiting values to define a number  
of subsamples. 
The Appendix discusses the solid and dashed lines; these show  
two Poisson cluster models that are able to provide reasonable  
descriptions of the measurements.   
 
\begin{figure*} 
 \centering 
 \includegraphics[width=\hsize]{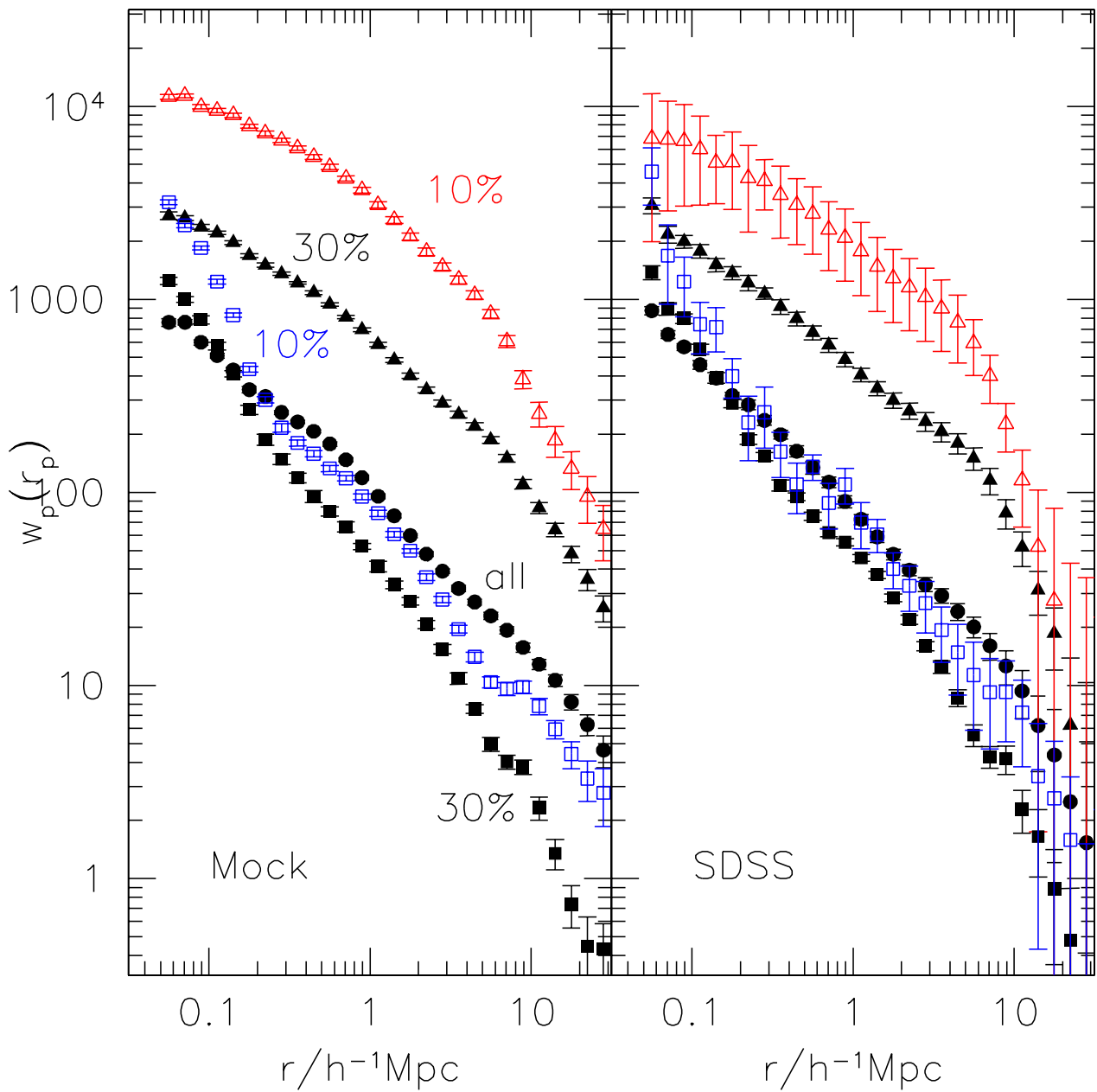} 
 \caption{Environmental dependence of clustering in the SDSS  
          for galaxies volume-limited to $M_r<-19.5$ (right)  
          and in a halo-model based mock catalog (left).   
          Filled circles show the clustering in the full sample;  
          filled and open triangles show subsets containing 30\%  
          and 10\% of the objects classified as being in the  
          densest regions; filled and open squares show similar  
          measurements but in the least dense regions.}   
 \label{xisdss} 
\end{figure*}

Filled circles in the right hand panel of Figure~\ref{xisdss}  
show the projected correlation function of the full sample;  
this is computed by integrating $\xi(r_p,\pi)$ over  
$0\le \pi \le 35h^{-1}$Mpc.   
Error bars are from jack-knife resampling in which the statistics  
were remeasured after omitting a random region, and repeated  
thirty times (approximately 1.5 times the total number of bins  
in separation for the results presented, as in Abbas \& Sheth 2006).   
Filled triangles show the corresponding measurement in the  
subsample which contains 30\% of the galaxies chosen to lie in  
the densest regions.   
Filled squares show the clustering in a sample of the same size,  
but now drawn from the least dense regions.   
The open triangles and squares show the result of selecting only  
the densest and least dense 10\%, rather than 30\%, of the sample.   
 
\begin{figure} 
 \centering 
 \includegraphics[width=\hsize]{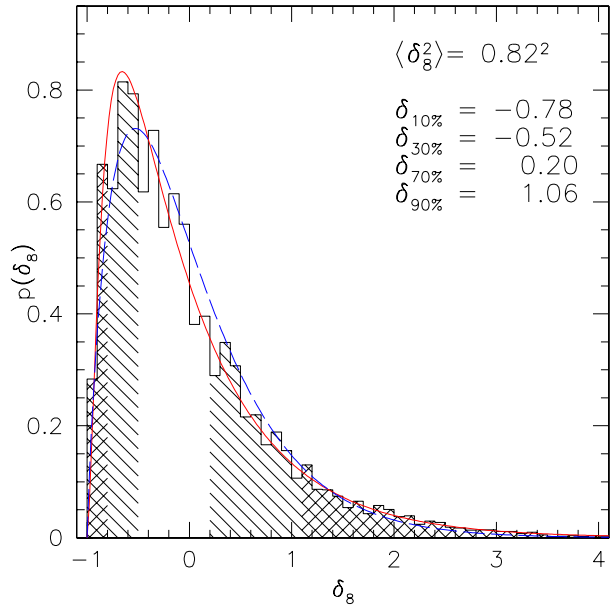} 
 \caption{Distribution of galaxy overdensity in a volume limited  
          catalog with $M_r<-19.5$ drawn from the SDSS.  The density  
          was defined by counting galaxies within $8h^{-1}$Mpc.   
          Dashed lines show where the area under the curve equals  
          10\%, 30\%, 70\% and 90\% of the total.  These correspond  
          to $\delta_8 =-0.78, -0.522, 0.196$, and 1.065.} 
 \label{sdsspdf} 
\end{figure} 
 
\begin{figure*} 
 \centering 
 \includegraphics[width=\hsize]{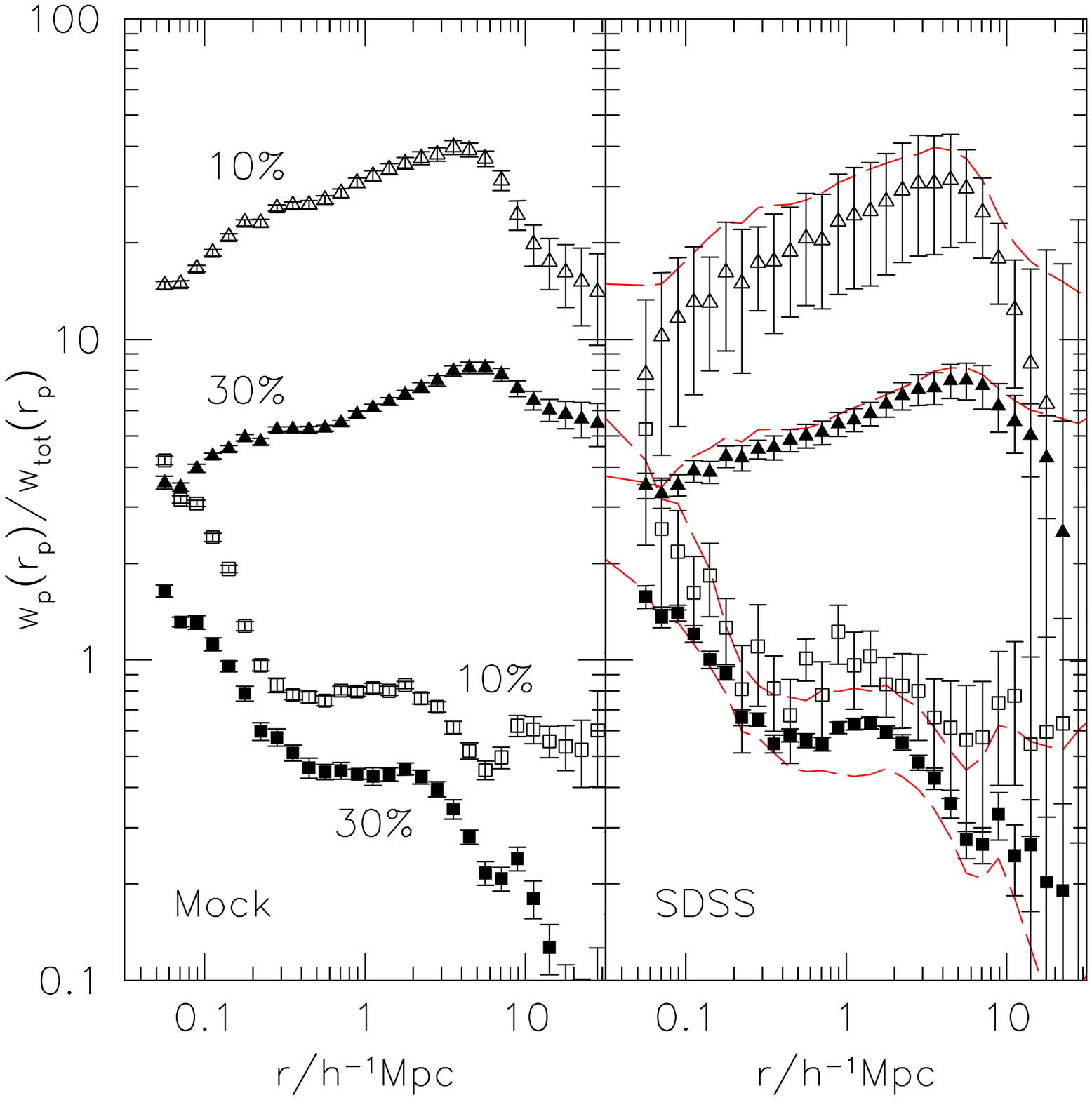} 
 \caption{Same as the previous figure, but now the clustering signal  
          is shown normalized to that for the full sample.  Dashed  
          lines in panel on the right show the locii traced out  
          by the points shown in the panel on the left.  } 
 \label{xiratio} 
\end{figure*} 
 
A number of unusual features are worth noting.   
First, on small scales, the correlation functions of all the  
subsamples have larger amplitudes than that of the parent sample  
from which they are drawn.   
This is most easily understood by supposing that the full sample  
is divided into two halves, say $D$ and $U$, for dense and  
underdense.  If the pair counts in the total sample are denoted  
TT, then the correlation function of the full sample is  
 $1+\xi_{tt}=TT/RR$,  
where $RR$ denotes the counts in an unclustered distribution of  
the same number density.  If we define $\xi_{dd}$ and $\xi_{uu}$  
similarly, then  
\begin{eqnarray} 
 1+\xi_{tt} &=& TT/RR = (DD + UU + 2DU)/RR \nonumber\\ 
            &=& (1+\xi_{dd})/4 + (1+\xi_{uu})/4 + 2(1+\xi_{du})/4.   
\end{eqnarray} 
However, on scales smaller than that on which the environment was  
defined ($8h^{-1}$Mpc in our case), $DU\approx 0$ by definition,  
so $\xi_{du}\approx -1$.  In this limit,  
 $\xi_{tt} = (\xi_{dd}+\xi_{uu})/4 - 1/2$, or  
\begin{equation} 
 \xi_{dd}+\xi_{uu} = 2(1 + 2\xi_{tt}). 
\end{equation} 
Thus, it is possible that $\xi_{dd}$ and $\xi_{uu}$ are both larger  
than $\xi_{tt}$. 
 
Second, on small scales, $\xi$ for the sample of galaxies in  
less dense regions can be substantially larger than it is for  
galaxies in denser regions.   
Abbas \& Sheth (2005) argue that this would arise if the average  
halo mass in underdense regions is smaller than it is in dense  
regions (see their Section 2.2).   
Evidence that this is the case comes from the fact that $\xi$  
shows a feature at $\sim 0.3h^{-1}$Mpc for the underdense  
sample, but not in the dense sample.  Abbas \& Sheth argue that  
this feature reflects the transition from pairs which are in the  
same halo to those which are in separate halos.  In the underdense  
regions, there are few neighbouring haloes within $8h^{-1}$Mpc (by  
definition), so this transition is obvious; since there may be many  
neighbouring halos in the dense regions, this transition is less  
obvious in the denser samples.   
 
A careful inspection suggests inflection points on scales of order  
$2h^{-1}$Mpc in the denser samples.  If this is due to the same  
transition, then the radii of halos in the denser samples are about  
$(2/0.3)$ times larger than in the least dense sample.   
It is standard to assume that the halos in dense and underdense  
regions have the same virial densities, so our measurements suggest  
that the halos in dense regions are typically about  
$(2/0.3)^3 = 300$ times more massive than those in the least dense  
sample. 
 
Finally, on larger scales where halo correlations are important,  
Figure~\ref{xisdss} shows that $\xi(r|\delta)$ is strongest in  
the densest regions.  This is not unexpected in the context of the  
linear peaks-bias model of (Kaiser 1984), if, on average, the densest  
regions at the present time formed from the densest regions in the  
initial fluctuation field.  This is because, in the initial Gaussian  
random field, the densest regions were more strongly clustered than  
regions of average density.  Although the galaxies in less dense  
regions are less strongly clustered than those in the very dense  
regions, the Figure suggests that the least dense 10\% are more  
strongly clustered than when the cut is 30\%.  Figure~\ref{xiratio},  
which shows the ratio of $\xi(r|\delta)/\xi(r)$,    
shows this effect slightly more clearly.  Although the  
difference on any given scale is only slightly larger than the  
error bars, it is in the same sense on all scales.  (While the  
errors on the measured $\xi$ are correlated between bins, the  
correlation between neighbouring bins is not expected to be strong,  
and it is expected to decrease with bin separation.  In any case,  
the next section shows that the effect is present with much  
larger statistical significance in mock catalogs of the effect.)   
 
\begin{figure*} 
 \centering 
 \includegraphics[width=0.8\hsize]{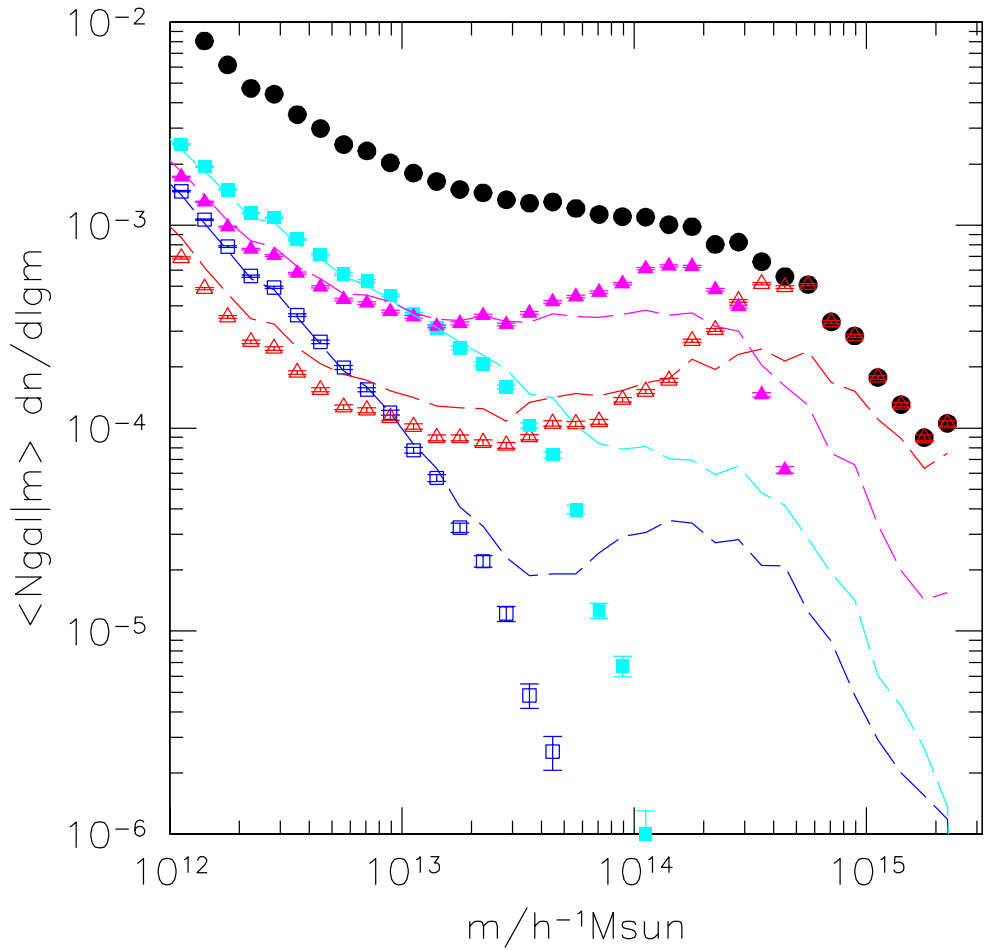} 
 \caption{Galaxy-weighted halo mass function as a function of  
          environment in our mock catalog.   
          Filled circles show this quantity when all  
          galaxies in the mock catalog are included.   
          The other sets of symbols show this quantity when  
          only galaxies in specific bins in environment are  
          used.  These bins are  
          the 10 percent of the objects with the fewest  
          neighbours within $8h^{-1}$Mpc ($N_8$),  
          the range between 10 and 30 percent,  
          the range between 70 and 90 percent,  
          and the 10 percent with the largest $N_8$.   
          Empty squares, filled squares, filled triangles  
          and empty triangles show results for these bins  
          when $N_8$ is defined using real space positions;  
          dashed curves show the corresponding measurement  
          when $N_8$ is in redshift space.   
          In real space, objects with the lowest $N_8$ values  
          populate the lowest mass halos.  While this remains true  
          in redshift space, there are a number of objects in  
          massive halos which appear to inhabit less dense regions;  
          these are galaxies in the Fingers of God of massive  
          halos which reach into less dense regions. } 
 \label{NgalMhalo} 
\end{figure*} 
 
\subsection{Measurements in mock galaxy samples}\label{mockgals} 
A more quantitative comparison between our model and the measurements  
is shown in the left hand panel of Figure~\ref{xisdss}.  The panel  
shows measurements of the environmental dependence of clustering in  
a mock galaxy catalog which was constructed as described by  
Abbas \& Sheth (2006).   
In brief, we assigned mock `galaxies' to halos in the VLS simulation  
by assuming that only halos more massive than a critical $m_{\rm L}$  
may contain galaxies.   
The first galaxy in a halo is called the `central' galaxy.   
The number of other `satellite' galaxies is drawn from a Poisson  
distribution with mean $N_s(m)$ where  
\begin{equation} 
 N_s(m) = (m/m_1)^\alpha \qquad  \text{ if }m \geq m_L.   
 \label{Ngsdss} 
\end{equation} 
We distribute the satellite galaxies in a halo around the halo centre  
so that the radial profile follows that of the dark matter (i.e., the  
galaxies are assumed to follow an NFW profile).   
We set $m_{\rm L}=10^{11.76}h^{-1}M_\odot$,  
 $m_1=10^{13.15}h^{-1}M_\odot$, and $\alpha=1.13$;  
Zehavi et al. (2005) show that these choices are appropriate for  
this set of SDSS galaxies: $M_r<-19.5$.  The resulting catalog has  
about $10^6$ mock galaxies, because the volume of our simulation  
is about ten times larger than that of our SDSS catalog.  This  
means that we can measure the environmental effects in the mock  
with greater precision than in the data.   
 
The important point, which we note explicitly here, is the following:   
By assuming that equation~(\ref{Ngsdss}) is the same function of $m$  
for all environments, and by assuming that the radial profile of the  
galaxies depends only on halo mass and not on environment, we have  
constructed a galaxy catalog in which all environmental effects are  
{\em entirely} a consequence of the correlation between halo mass and  
environment.  Therefore, the locii traced out by the various sets of  
symbols shown in Figure~\ref{xisdss} represent the predicted  
environmental dependence of $\xi$ if there are no environmental  
effects other than the statistical one determined by the initial  
fluctuation field.  The measurements in the mock catalog are extremely  
similar to those in the SDSS itself, leaving little room for  
additional environmental effects.   
 
In the mock catalog, the halo mass function in dense regions is  
top-heavy.  This is illustrated in Figure~\ref{NgalMhalo}, which  
shows the abundance of halos per logarithmic bin in mass, weighted  
by the number of galaxies in the mass bin.  Symbols show this  
galaxy-weighted halo mass function for bins in environment, where   
the environment is defined by the number of neighbours $N_8$  
in real space.  Curves show the corresponding measurement when  
$N_8$ is defined (as it is in the SDSS data) from redshift space  
positions.  Although Fingers of God tend to scatter some galaxies  
in massive halos into less dense environments, notice that the  
mix of halos is shifted towards lower masses in the less dense  
regions.   
 
The precise mix of halos determines the amplitude of the correlation  
function on both large and small scales, so the agreement seen on  
all scales in Figures~\ref{xisdss} and~\ref{xiratio} suggests that  
the halo abundances shown in Figure~\ref{NgalMhalo} are representative  
of those in the SDSS:  massive halos preferentially populate  
dense regions.  Indeed, our estimate of a factor of $\sim 100$  
difference in mass (based on Figure~\ref{xisdss}) appears to be  
in good agreement with the mass functions shown in  
Figure~\ref{NgalMhalo}.   
 
The fact that the mock catalog accurately reproduces the inflection  
in $\xi(r|\delta)$ seen at $\sim 0.3h^{-1}$Mpc in the underdense  
regions has an interesting implication.   
Abbas \& Sheth (2005) show that this inflection scale reflects the  
typical virial diameters of halos in these environments.  
Therefore, the agreement with the SDSS suggests that the mock catalog  
has modeled the correlation between halo mass and virial radius  
accurately.  Since modelers differ in what this density should be  
(200 times background density? 200 times critical density? some  
other multiple of background density?), the agreement is nontrivial.   
In the mock, halos are 200 times the critical density whatever  
their mass.  If they were 200 times the background density instead,  
they would be larger by a factor of $\Omega_0^{-1/3}\approx 3/2$.   
As samples get larger, $\xi(r|\delta)$ will become more precisely  
measured, and so it may provide an interesting constraint on halo  
densities.   
 
\section{Linear bias and environmental dependence of clustering}\label{nlbias} 
The previous section showed that the large scale clustering  
strength is not a monotonic function of environment.   
A simple model of this effect follows from writing the linear  
peaks bias model in terms of the nonlinear density $\delta$:   
\begin{equation} 
 \xi(r|\delta) \approx  
   \left[{\delta_0(\delta)\over \sigma_\delta}\right]^2\, 
    {\xi(r)\over \sigma_\delta^2} = B_\delta^2\,\xi(r) 
\end{equation} 
(equation~5 of Kaiser 1984 with Sheth 1998b,c), 
where $\delta_0(\delta)$ denotes the value of the density contrast  
in linear theory when the fully nonlinear overdensity is $\delta$,  
and $\sigma_\delta$ denotes the rms value of the linear  
fluctuation field when smoothed on a scale which contains mass  
$\bar\rho V(1+\delta)$.  For a power-law power spectrum, 
 $\sigma_\delta^2 = \sigma_0^2/(1+\delta)^{(n+3)/3}$  
so  
\begin{equation} 
 B_\delta = {\delta_c\,[1 - (1+\delta)^{-1/\delta_c}]\over  
             \sigma_0^2\,(1+\delta)^{-(n+3)/3}}. 
 \label{Bdelta} 
\end{equation} 
where our relation between $\delta_0$ and $\delta$ provides an  
excellent description of the spherical collapse model if we set  
$\delta_c\approx 1.686$.   
Note that, in our case, $\sigma_0=\sigma_8=0.9$.   
 
\begin{figure} 
 \centering 
 \includegraphics[width=\hsize]{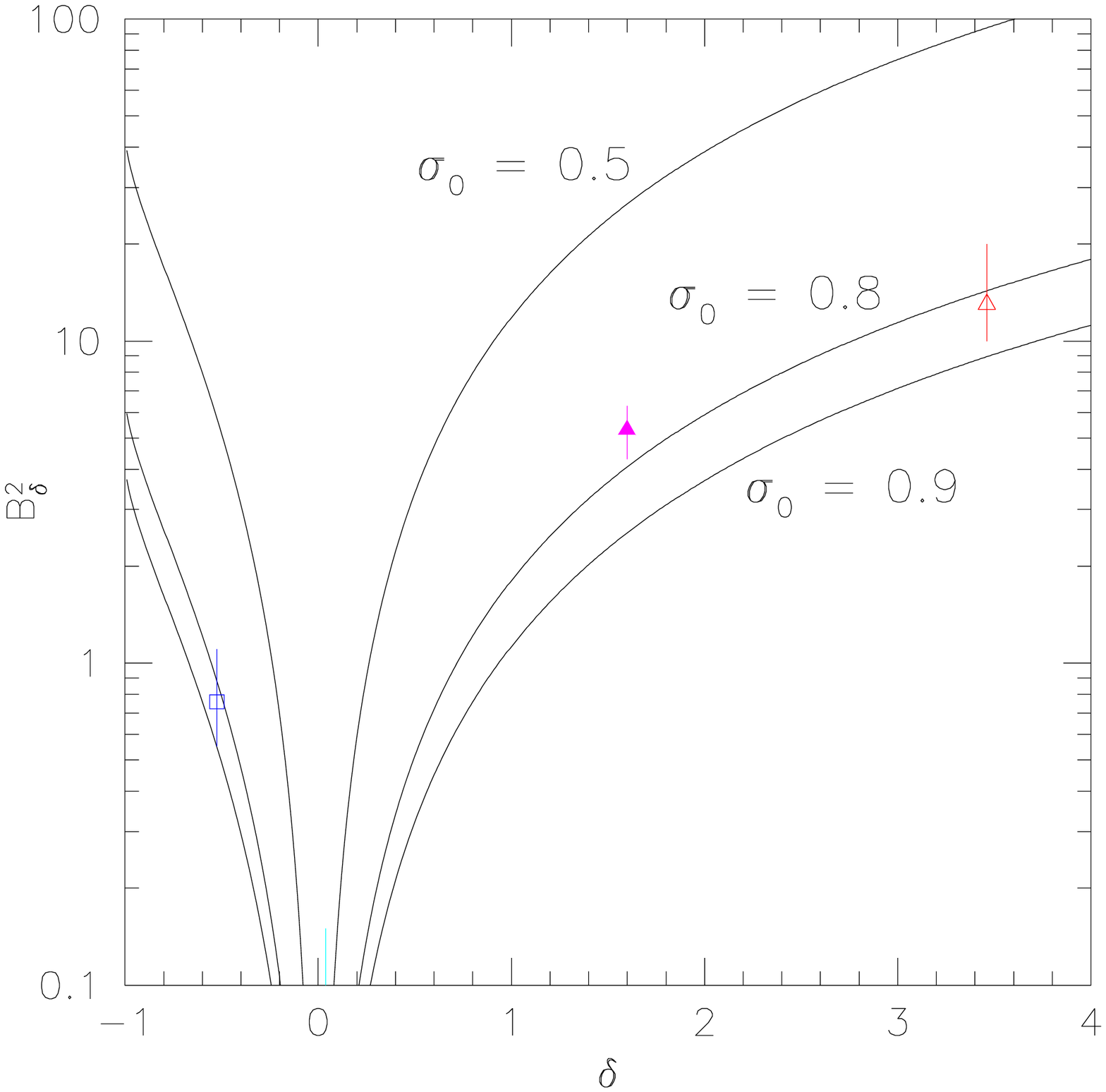} 
 \caption{Bias as a function of environment.  Curves show  
          equation~(\ref{Bdelta}) with $n=-1.2$ and three values  
          of $\sigma_0$:  $B_\delta^2$ is symmetric about $\delta=0$  
          for small values of $\sigma_0$, but becomes increasingly  
          asymmetric as $\sigma_0$ increases.   
          Symbols (same as in Figure~\ref{xiratio})  
          show the bias as a function of environment derived from  
          Figure~\ref{xiratio} following the procedure described  
          in the text.  } 
 \label{toybias} 
\end{figure} 
 
Equation~(\ref{Bdelta}) shows that $B_\delta^2$ is not a monotonic  
function of $\delta$, nor is it symmetric around $\delta=0$.   
Figure~\ref{toybias} shows this explicitly for $n=-1.2$ (since  
this would produce a correlation function with slope $-1.8$,  
which is approximately the slope we see for the full sample)  
and a few values of $\sigma_0$ (appropriate for a scale of  
about $8h^{-1}$Mpc).  While this simple model is qualitatively  
consistent with our measurements, making a more quantitative  
statement is less straightforward.   
 
The symbols show a very rough indication of the how the bias  
scales with environment in the previous figures.  We caution that  
the locations of these points are {\em not} model free because  
$\delta$ in the expression above is for the dark matter, whereas  
our environment is defined by the galaxy distribution.   
To place the points on this plot, we have assumed that the galaxies  
in the total sample are unbiased (so $\delta = \delta_{\rm gal}$  
and $\xi_{\rm all}$ (the filled circles in Figure~\ref{xisdss})  
equals the correlation function of the dark matter which appears in  
the right hand side of equation~(\ref{Bdelta}).   
For each value of $\delta_8$ in Figure~\ref{sdsspdf} we determined  
a $\delta_{\rm gal}$ from requiring that  
 $\bar N[1+\delta_{\rm gal}] = (1+\delta_8)[1 + \bar N(1+\bar\xi)]$;  
we used $\bar N = 0.01\,4\pi 8^3/3 = 21.45$ and  
$\bar N \bar\xi = 1/0.2^2 - 1 = 24$, as suggested by the fact  
that equation~(\ref{gp}) with $b=0.8$, when inserted in  
equation~(\ref{qN}), provides a good description of the distribution  
of $N_8$ in the data (c.f. solid line in Figure~\ref{sdsspdf}).   
This procedure accounts for the fact that $\delta_8$ is computed in  
cells centred on galaxies, whereas $\delta_{\rm gal}$ is not.   
(Equation~\ref{Nbarq} and associated discussion shows why this  
scaling is reasonable.)  The associated bias factors for these  
points were read-off from the large scale values of the ratio  
shown in Figure~\ref{xiratio} (recall we are assuming that the  
full sample is unbiased relative to the dark matter).

While the agreement is reassuring (e.g. this procedure correctly  
predicts {\em very} weak clustering for the sample which contains  
the 30\% of the objects with lowest $N_8$), a better analysis of  
the effects of bias is required to make this model more than simply  
illustrative.  Our main point in the present work is to show that  
the clustering strength is not expected to be a monotonic function  
of environment.   
 
\section{Discussion and Conclusions} 
High peaks and low troughs in a Gaussian random field are similarly  
biased relative to the set of all fluctuations.  Although nonlinear  
evolution destroys this symmetry, some remnants of it are expected to  
remain in the galaxy clustering signal (equation~\ref{Bdelta}  
and~Figure~\ref{toybias}).  We find that galaxies in less dense  
regions of the SDSS are less strongly clustered than galaxies in  
very dense regions, but that galaxies in the least dense regions  
are more strongly clustered than galaxies in regions of moderate  
underdensity (Figures~\ref{xisdss} and~\ref{xiratio}).   
 
Our simple model for this effect (equation~\ref{Bdelta}) makes a  
qualitative prediction which can soon be tested:  namely, the  
strong clustering of underdense regions is more easily noticed  
when $\sigma_0\ll 1$.  This is because $B_\delta^2$ is more  
symmetric about $\delta=0$ for small values of $\sigma_0^2$:   
Figure~\ref{toybias} shows this clearly.  At fixed redshift,  
this means that the effect should be easier to notice for  
environments defined on larger scales.  Alternatively, for  
environments defined on a fixed comoving scale, the fact that  
clustering is not a monotonic function of environment should be  
easier to notice at higher redshift.   
In our analysis of the SDSS, we had to go to extremely underdense  
environments before we found evidence that clustering was not  
monotonic with environment.  Our model suggests that, at  
higher redshift, one need not go to as extreme values of $\delta$  
to see the enhanced clustering of underdense regions.   
 
We also showed that a mock galaxy catalog, constructed to reproduce  
the clustering of the full sample, exhibits the same environmental  
dependent clustering signals as seen in the SDSS (Figures~\ref{xisdss}  
and~\ref{xiratio}).  This agreement is non-trivial---rather than  
being simple power laws with different amplitudes, the correlation  
functions in different environments exhibit inflections on various  
different scales.  The agreement suggests that a number of features  
of the mock are also true of the Universe:  namely, the mix of  
dark matter halo masses is top-heavy in dense regions  
(Figure~\ref{NgalMhalo}), massive halos have larger virial radii,  
and the primary drivers of correlations between galaxy luminosity  
and environment are the correlations between galaxy luminosity and  
host halo mass, and the correlation between halo mass and environment.   
In particular, there is little room for the additional effects which  
Sheth \& Tormen (2004) show may also play a role in determining these  
correlations (also see Gao et al. 2005; Harker et al. 2005;  
Wechsler et al. 2006; Zhu et al. 2006).   
 
Environmental effects are also present in Poisson cluster models  
(Appendix).  In such models, halo bias is simply a consequence of  
mass conservation:  whereas the origin of halo bias is usually  
stated as arising from the fact that dense regions host massive  
halos, in Poisson cluster models, dense regions are dense precisely  
because they happen to host dense halos.   
Nevertheless, these models predict halo bias relations which are  
surprisingly like those in the standard model of halo bias  
(Mo \& White 1996; Sheth \& Tormen 2002).  The analysis in the  
Appendix is particularly interesting in view of the fact that two  
Poisson cluster models, the Thermodynamic or Generalized Poisson  
(Saslaw \& Hamilton; Sheth 1995) distribution and the Negative  
Binomial distribution, both provide good descriptions of our data  
(Figure~\ref{sdsspdf}, and also see recent analyses of the void  
probability function by Croton et al. 2004 and Conroy et al. 2005).   
 
All other cosmological parameters remaining fixed, larger values  
of the rms fluctuation amplitude $\sigma_8$ imply a larger range of  
environments.  So the difference between the densest and least dense  
regions increases with increasing $\sigma_8$.  Therefore, one might  
expect the environmental dependence of clustering to yield useful  
information about $\sigma_8$, although the results in  
Tinker et al. (2006) suggest otherwise.  So it is interesting  
that the dashed lines in Figure~\ref{xiratio} tend to lie slightly  
further from unity than do the SDSS measurements.   
Determining whether or not this is indicating that the data prefer  
a value of $\sigma_8$ which is lower than the value ($\sigma_8=0.9$)  
used in the mocks is the subject of work in progress.   
 
\section*{Acknowledgements} 
We thank Jeremy Tinker and Idit Zehavi for discussing the existence  
of this effect in their mocks, the referee P. Monaco for a helpful  
report, the Virgo consortium for making their simulations available  
to the public, Ryan Scranton for providing the SDSS data, Jeff Gardner  
for providing the Ntropy code which was used to measure the projected  
correlation functions in the simulations and the data, and the Aspen  
Center for Physics for hospitality while this work was completed.   
This work was supported in part by NSF-AST 0520647.

\appendix 
 
\section{Environmental effects in Poisson cluster models}\label{envpcp} 
The main text argued that the correlation function depends on  
environment primarily because the halo distribution is top-heavy  
in dense regions.   
While it is tempting to conclude that this derives from the fact  
that more massive halos are more strongly clustered, this is not  
the whole story.  The following calculation illustrates that, even  
in Poisson cluster models, the distribution of halos depends on  
environment.  This is a simple consequence of mass conservation:   
a region containing $N$ particles may not host a halo with mass $n>N$.   
 
As the name suggests, Poisson cluster models are point  
distributions in which cluster centers are distributed at random  
(i.e., the distribution of clusters is Poisson); different models  
are distinguished by specifying the probability that a randomly  
selected cluster contains $n$ galaxies.  The clusters themselves  
are assumed to have zero size.  Poisson cluster distributions  
are also sometimes called Compound Poisson (Daley \& Vere-Jones 2003).   
 
\subsection{Counts-in-cells for unclustered clusters}\label{compois} 
Let $p(N|V)$ denote the probability that a randomly placed cell  
(i.e. not necessarily centred on a particle) of volume $V$ contains  
$N$ particles, and define  
\begin{equation} 
 P(s|V) \equiv \sum_N s^N\, p(N|V).   
\end{equation} 
For a Compound Poisson distribution,  
\begin{equation} 
 \ln P(s|V) = \bar N \Bigl[H(s) - 1\Bigr] 
                    \Bigl[\partial H(s)/\partial s|_{s=1}\Bigr]^{-1}, 
 \label{cpgf} 
\end{equation} 
where $\bar N$ is the mean of $p(N|V)$,  
\begin{equation} 
 H(s) \equiv \sum_n s^n h(n),  
\end{equation} 
and $h(n)$ denotes the probability that a randomly chosen cluster  
contains $n$ particles (e.g. Daley \& Vere Jones 2003).   
 
In the main text we defined the environment of a particle (in  
that case a galaxy) by counting the number of particles in a cell  
of volume $V$ centred on it.  Hence, we seek an expression for  
the probability $q(N|V)$ that a cell of volume $V$, centred on a  
randomly chosen particle, also contains $N-1$ other particles.   
For Compound Poisson distributions, this distribution is simply  
related to $p(N|V)$, the distribution of counts in randomly placed  
cells.  This is because  
\begin{equation} 
 q(N|V) \equiv {\sum_{n=1}^N nh(n)\, p(N-n|V)\over \sum_{n>0} nh(n)}; 
\end{equation} 
the terms involving $h(n)$ denote the probability that $V$ is centred  
on a particle in an $n$ halo, and the term involving $p$ denotes the  
probability that $V$ contains $N-n$ particles in addition to the $n$  
which are associated with the halo in which the chosen particle sits.   
Note that the term in the denominator above is $\partial H(s)/\partial s$  
evaluated at $s=1$.  In what follows, we will denote this quantity  
$H'(s=1)$.   
 
To derive an expression for $q$, it is convenient to begin with  
\begin{eqnarray} 
 Q(s|V) &\equiv& \sum_{N>0} s^N\,q(N|V) \nonumber\\ 
   &=& \sum_{N>0} s^N \sum_{n=1}^N {nh(n)\,p(N-n|V)\over H'(s=1)}\nonumber\\ 
   &=& \sum_{N>0} \sum_{n=1}^N  
         {s^n\,nh(n) s^{N-n}\,p(N-n|V)\over H'(s=1)}\nonumber\\ 
   &=& \sum_{n>0} {s^n\,nh(n)\over H'(s=1)}  
         \sum_{N>n-1} s^{N-n}\,p(N-n|V)\nonumber\\ 
 % &=& \sum_{n>0}{s^n\,nh(n)\over H'(s=1)}\sum_{M>0}s^{M}\,p(M|V)\nonumber\\ 
   &=& \sum_{n>0} {s^n\,nh(n)\over H'(s=1)} P(s|V)  
    =  s\, {H'(s)\over H'(s=1)} \, P(s|V)\nonumber\\ 
   &=& {s\over \bar N} {\partial P(s|V)\over \partial s} 
    = {s\over \bar N} {\partial \over \partial s}\sum_N s^N p(N|V)\nonumber\\ 
   &=& {s\over \bar N} \sum_{N>0} N s^{N-1} p(N|V).   
 \label{Qs} 
\end{eqnarray} 
Comparison of the first and last expressions shows that  
\begin{equation} 
 q(N|V) = {N\over\bar N}\,p(N|V).   
 \label{qN} 
\end{equation} 
Evidently, if one only considers volumes which are centred on  
particles, then the distribution of counts in such cells is simply  
related to the distribution of counts in randomly placed volumes---the  
two distributions differ by one factor of $N/\bar N$.   
 
This useful result follows from two assumptions:   
clusters have vanishingly small sizes, and they are uncorrelated  
with one-another. 
So long as we restrict attention to volumes $V$ which are large  
compared to the virial radius of a typical cluster (about 2~Mpc),  
the assumption that halos have negligible sizes should be  
reasonable.  The neglect of cluster clustering in our Universe  
is less reasonable, but, as we show below, is still a useful  
approximation.   
 
Equation~(\ref{Qs}) implies that $\bar N_q$, the mean of $q$ is  
\begin{equation} 
 \bar N_q \equiv {\langle N\rangle\over \bar N}  
                 + {\langle N(N-1)\rangle\over \bar N}  
          = 1 + \bar N\,(1+\bar\xi) 
 \label{Nbarq} 
\end{equation} 
where $\bar\xi$ is the volume average of the two point correlation  
function.  This makes intuitive sense:  the mean count when  
centred on a particle is one plus $\bar N$, plus a contribution  
from the fact that the particles are correlated.

\subsection{The Generalized Poisson and Negative Binomial  
            distributions}\label{gpd} 
In what follows, we use the cluster mass function associated with an  
initially Poisson distribution:   
\begin{equation} 
 h(n) = {(nb)^{n-1}\exp(-nb)\over n!},\ \ {\rm where}\ \ b=(1+\delta_c)^{-1} 
 \label{borel} 
\end{equation} 
(Epstein 1983; Sheth 1995).   
Here $\delta_c$ is the critical density in the initial density  
fluctuation field that is required for collapse in the spherical  
model.  Thus, $b=0$ initially, and it grows to $b\to 1$.   
% Note that $\sum_n n\,h(n) = (1-b)^{-1}$.   
If these clusters have a Poisson distribution, then  
\begin{eqnarray} 
 p(N|V) &=& {\bar N(1-b)\over N!}\,  
            \left[\bar N(1-b) + Nb\right]^{N-1}\nonumber\\ 
         && \qquad\times\quad \exp[-\bar N(1-b) - Nb] 
 \label{gp} 
\end{eqnarray} 
is the Generalized Poisson distribution (e.g. Sheth 1998b).   
Here $\bar N$ is the average number of particles in a cell  
of size $V$:  $\langle N\rangle=\bar N$.   
This distribution, which is sometimes called the Thermodynamic  
distribution in the astrophysical literature (Saslaw \& Hamilton 1984;  
Sheth 1995), provides a reasonably good description of the counts  
of galaxies in randomly placed cells of size $V$  
(Hamilton, Saslaw \& Thuan 1985; Sheth, Mo \& Saslaw 1994;  
Conroy et al. 2006).   
 
For this distribution, the variance is   
 $\langle N^2\rangle - \langle N\rangle^2 = \bar N/(1-b)^2$ 
so the mean value of $q$ is $\bar N + 1/(1-b)^2$.   
The solid line in Figure~\ref{sdsspdf} shows $q(N|V)$ associated  
with equation~(\ref{gp}), where $\bar N = 0.01\,4\pi 8^3/3 = 21.45$  
and $b=0.8$.  It provides a good description of the measured  
distribution.   
 
The dashed line shows a similar analysis of the Negative Binomial  
distribution.  This distribution has  
\begin{equation} 
 p(N|V) = {(\gamma + N-1)!\over N!\,(\gamma-1)!}\, \beta^{N}\, (1-\beta)^\gamma 
 \label{NegBin} 
\end{equation} 
with $\gamma \equiv \bar N\, (1-\beta)/\beta$.   
The Negative Binomial is a Compound Poisson distribution with  
\begin{equation} 
 h(n) = {\beta^n/n\over -\ln(1-\beta)}.   
 \label{logarithmic} 
\end{equation} 
The mean and variance of this distribution are $\bar N$ and  
$\bar N/(1-\beta)$, so we have set $\bar N = 21.45$ as before,  
and $\beta = 1 - (1-b)^2 = 0.96$ (so the variance also matches  
that of the Generalized Poisson distribution and the data).   
 
Figure~\ref{sdsspdf} suggests that the Generalized Poisson  
distribution provides a slightly better description of this  
dataset than does the Negative Binomial.  We are not as  
interested in which provides a better fit, as we are in the  
fact that a Compound Poisson model appears to work so well.   
 
\subsection{Halo bias in Compound Poisson distributions}\label{cpbias} 
Having made the connection between counts-in-cells and the cluster  
distribution, we now consider environmental effects in Poisson cluster  
models.   
The mean density of $n$-halos which are surrounded by regions which  
contain $N$ particles on the scale $V$ is  
\begin{equation} 
 h(n|N) = {\bar n h(n)\over\sum nh(n)}\, {p(N-n|V)\over p(N|V)}. 
\end{equation} 
The ratio of this to the average density of $n$-halos is  
$p(N-n|V)/p(N|V)$.  Since this ratio is obviously different from  
unity, the mass function of halos in dense regions is different from  
underdense regions.  Typically, $p(N|V)$ drops exponentially when  
$N\gg\bar N$:  $p(N|V) \propto \exp(-\alpha N)$ with $\alpha > 0$.   
Thus, $p(N-n)/p(N)\approx \exp(\alpha n)$:  massive halos are  
exponentially more abundant in regions with large $N$.   
 
The following explicit calculation shows that the Poisson cluster  
model actually captures much of the usual parametrization of the  
environmental dependence of halo abundances.  That is, the following  
demonstrates that mass conservation itself provides a significant  
source of halo bias.  Since mass conservation is most important  
on scales $V$ where the typical mass in a cell is not substantially  
larger than the typical halo mass, we expect the Poisson cluster model  
to provide a reasonable approximation on such scales.   
(And recall that Figure~\ref{sdsspdf} shows that the  
Generalized Poisson and Negative Binomial distributions do indeed  
provide good descriptions of the data.)   
 
\subsection{Halo bias in the Generalized Poisson distribution}\label{gpdbias} 
If we define $NB\equiv \bar N(1-b) + Nb$ (the reason for this notation  
will become clear shortly), then the density of $n$ halos surrounded  
by regions of size $V$ which contain $N$ particles is 
\begin{eqnarray} 
 h(n|N)  
% &=& {\bar n(1-b)\over \bar N(1-b) + Nb}\,{N\choose n}\, 
%          \left({nb\over \bar N(1-b) + Nb}\right)^{n-1}\,\nonumber\\ 
%  &&\quad\times\quad  
%          \left({\bar N(1-b) + (N-n)b\over\bar N(1-b) + Nb}\right)^{N-n-1} 
  &=& \left({B-b\over B\,V}\right) {N\choose n} 
          \left({nb\over NB}\right)^{n-1}\! 
          \left(1 - {nb\over NB}\right)^{N-n-1} 
\end{eqnarray} 
where we have used the fact that  
$1-b/B = (NB - Nb)/NB = \bar N(1-b)/[\bar N(1-b) + Nb]$. 
This form is precisely that of the conditional mass function of  
$n$ halos in the Poisson model (Sheth 1995, 2003).   
This is remarkable for the following reason.   
 
The usual estimate of the environmental dependence of halo abundances  
uses the conditional mass function, and it uses the spherical evolution  
model to transform the density $N/\bar N$ to an initial overdensity  
(Mo \& White 1996; Sheth \& Tormen 2002).   
To see what the transformation is in the present case,  
set $B=(1+\delta_0)^{-1}$ (this is motivated by the fact that  
$b=(1+\delta_c)^{-1}$.   
Then $B = \bar N(1-b)/N + b$, so  
$1+\delta_0 = (1+\delta_c)(1+\delta)/[\delta_c + 1+\delta]$.   
Hence, in a model in which the halos in the unconditional mass function  
are assumed to have a Poisson spatial distribution, the mapping between  
linear and nonlinear density is  
\begin{equation} 
 \delta_0 = {\delta_c\over 1 + \delta_c/(1+\delta)}\,{\delta\over 1+\delta} 
          \approx \delta_c\,{\delta\over 1+\delta}. 
\end{equation}   
This relation is qualitatively like that of the spherical model,  
which is very well approximated by  
\begin{equation} 
 1+\delta \approx (1 - \delta_0/\delta_c)^{-\delta_c}. 
\end{equation} 
See Sheth (1998b,c) for more discussion of the similarities between this  
and the spherical model, and another derivation of the relation between  
equations~(\ref{borel}) and~(\ref{gp}).   
This qualitative similarity, and the fact that the Generalized Poisson  
model appears to describe the data in Figure~\ref{sdsspdf} reasonably  
well, both suggest that Poisson cluster models may provide useful  
insight into the origin of environmental effects.   
 
In these models, environmental effects arise not because dense  
regions host the most massive halos, but because a region which  
contains a massive halo tends to be denser than average, simply  
because of mass conservation.   
This is a rather different view of environmental effects than that  
of Mo \& White (1996) and Sheth \& Tormen (2002), where the large  
scale environment, rather than mass conservation, is seen as the  
primary driver of the correlation between halo mass and environment!   
 
The formation histories of halos in Poisson cluster models have  
been studied in Sheth (1998a).  By combining that analysis with  
the present one, it should be interesting and straightforward to see  
what correlations between formation history and environment are built  
into such models.   In addition, it would also be interesting to  
repeat this analysis for the Negative Binomial distribution.   
 
\label{lastpage} 
 
\end{document}